# Nanomechanical characterization of the fracture toughness of Al/SiC nanolaminates

L. W. Yang[a,b], C. R. Mayer[c], N. Chawla[c,d], J. Llorca[a, e,*], J.M. Molina-Aldareguía[a,*]

[a]IMDEA Materials Institute, C/Eric Kandel 2, 28906, Getafe, Madrid, Spain

[b]Hypervelocity Aerodynamics Institute, China Aerodynamics Research and Development Center, 621000, Mianyang, China

[c]Center for 4D Materials Science, Arizona State University, Tempe, Arizona, AZ 85287, USA

[d]School of Materials Engineering, Purdue University, West Lafayette, IN 47907,

[e]Department of Materials Science, Polytechnic University of Madrid, E.T.S. de Ingenieros de Caminos, 28040 Madrid, Spain

**Abstract**

The fracture toughness of Al/SiC nanolaminates with different layer thicknesses (in the range 10 to 100 nm) was measured by means of micropillar splitting and bending of a notched beam. The crack plane was perpendicular to the layers in the former while notched beams with the notch parallel and perpendicular to the layers were milled in the latter. It was found that crack propagation parallel to the layers took place along the metal-ceramic interfaces and the toughness increased with the layer thickness due to the contribution of the plastic deformation of the Al layers. Crack propagation perpendicular to the layers showed evidence of crack deflection/arrest at the interface. The toughness in this orientation increased as the layer thickness decreased due to the higher density of interfaces except for the nanolaminates with 10 nm layer thickness. In the latter case, crack propagation took place along the weak columnar grain boundaries, leading to a marked reduction in toughness.

# 1. Introduction

Nanolaminates made up of alternating metallic and ceramic nanoscale thick layers are promising materials in numerous engineering applications, due to their outstanding mechanical properties including high strength and wear resistance [1–6], as well as unique electrical and optical properties [7-8]. It has been argued that the fracture toughness of metal-ceramics nanolaminates can be higher than that of conventional metal-ceramic composites due to several reasons. Firstly, cracking of the brittle ceramic layers can be delayed due their nanoscale dimensions, which limit the size of the pre-existing flaws that initiate fracture. Secondly, the energy dissipated by the plastic deformation of metallic layers can arrest cracks at the metal-ceramic interfaces. And finally, the large number of interfaces associated with the nanoscale dimensions of the layers can introduce substantial energy dissipation through crack deflection mechanisms [9].

Nevertheless, there is very little reliable information on the toughness properties of nanolaminates because most of them are fabricated in the form of thin-films or coatings, and appropriate techniques to measure the fracture toughness of coatings are still lacking [10-12]. Several techniques have been proposed to measure fracture toughness at the microscale, most of them based on bending of cantilevers of different geometries, such as Chevron notch cantilevers [13,14], clamped beams [15] or double cantilever beams [16-18]. These approaches are very time consuming (because they require the fabrication of the beams by focused ion beam milling) and ion-induced damage at the root of the pre-notch may introduce additional artefacts [19]. Another approach is based on the micropillar splitting method, developed by Sebastiani *et al.* [20, 21], which does not require the introduction of pre-notches. Instead, the cracks are introduced directly by a sharp pyramidal indenter and propagate in mode I.

In this investigation, both micropillar splitting and bending of notched cantilevers were used to determine the fracture toughness of Al/SiC nanolaminates with different layer thicknesses (in the range 10 nm to 100 nm).

# 2. Materials and experimental techniques

The Al/SiC nanolaminates were fabricated by magnetron sputtering physical vapor deposition on a Si wafer in Los Alamos National Laboratory. The sputter unit is made up of dual sputter guns for the deposition of Al and SiC in a high vacuum chamber, using high purity Al and SiC targets (>99.5%, Kurt J. Lesker, Clairton, PA). The designed multilayer structure of Al/SiC was built up by means of a computer controlled shutter system. The deposition rates were ~7.5 nm/min for Al and ~3.9 nm/min for SiC. 4 Al/SiC nanolaminates with the same nominal volume fraction of Al (50%) and SiC(50%) and nominal individual layer thicknesses of 10, 25, 50 and 100 nm were fabricated. They were labelled AS10, AS25, AS50 and AS100 in this work, respectively. The microstructure of the AS10 and AS100 nanolaminates was characterized by transmission electron microscopy (TEM) in a FEI Talos$^{TM}$ F200 system. The Al layers were nanocrystalline, with the vertical grain size limited by the layer thickness, while the SiC layers were amorphous. The coatings displayed a columnar microstructure, typical of magnetron sputtered thin-films, are shown in Fig. 1 (a-b). Detailed information about the layer thickness, the Al grain size and the columnar width is summarised in Table 1. More details about the microstructure can be found in [22]. The residual stresses in the Al/SiC nanolaminates were determined by X-ray diffraction [23]. They were of the order of -50 MPa in the Al layer and no significant variation of residual stresses with layer thickness was found. The nanolaminate was magnetron sputtered on Si wafers at room temperature and the residual stresses were attributed to the peening mechanism due to bombardment of the Al layers by SiC and Ar neutrals during deposition.

The fracture toughness of the Al/SiC nanolaminates was measured using two different techniques, namely micropillar splitting [20-21] and fracture of notched cantilever beams [13-18].

Micropillar splitting involves loading a circular micropillar by a sharp cube-corner indenter up to a load that induces the nucleation and propagation of cracks from the corners of the triangular indentation imprint and the splitting fracture of the micropillar. Micropillars of ~3 µm in diameter and aspect ratio ~1 were manufactured by conventional focused ion beam milling (FIB) in a dual beam scanning electron microscope (SEM) (Helios Nanolab 600i FEI) (Fig. 1c). The pillar axis was perpendicular to the layers to ensure symmetric cracking during splitting. The splitting tests were carried out using the Hysitron Triboindenter$^{TM}$ system, which allows positioning of the tip at the center of the micropillar top surface with a precision better than 10 nm, preserving the symmetry during the test.

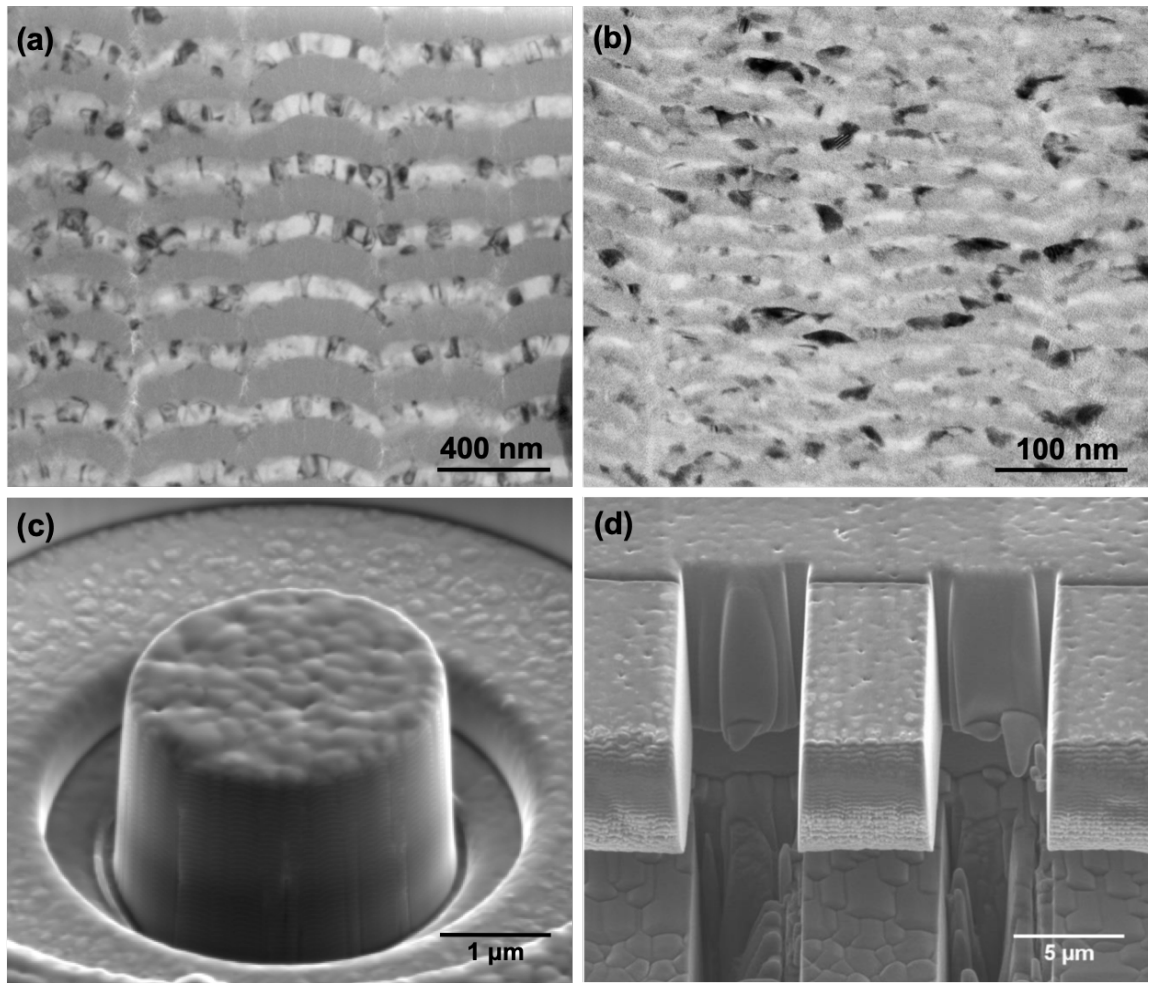

Fig. 1. Typical TEM microstructures of Al/SiC nanolaminate for (a) AS100 and (b) AS10 nanolaminates. (c) Micropillar of 3 µm in diameter and aspect ratio ≈ 1 of the AS25 nanolaminate used in the micropillar splitting tests. (d) As-milled notched beam (dimensions: 5µm × 5µm × 13µm) of the AS100 nanolaminate.

Manufacturing of the cantilever beams requires a polished 90° edge to be exposed. To achieve this, approximately 5 mm by 5 mm sections of wafer were adhered to a 1 × 1 cm$^2$ copper block with a small amount of wafer overhanging the copper block. This overhanging section was then polished away flush to the block, providing a 90° corner with which to work. The cantilever beams were fabricated using a dual beam SEM (Nova 200, FEI) with an ion beam accelerating voltage of 30 kV. First, a high current of 20 nA was used to mill a large trench in the side of the sample, leaving an approximately 12 µm × 60 µm × 7 µm (depth × length × thickness) free-standing foil. Then, using a 7 nA current normal to the top of the sample, this foil was divided into 5 individual cantilevers, approximately 7 µm wide. The shape of the cantilevers was then refined in steps to their nominal 5 µm × 5 µm × 13 µm dimensions using decreasing currents down to 0.1 nA. Line milling at 0.1 nA was used to cut the notches, whose depths ranged from 500 - 1000 nm because different samples and orientations have slightly variable milling rates. Five beams were fabricated for each combination of material (AS10, AS50 and AS100) and orientation (layers parallel and perpendicular to the notch plane). A SEM image of a finished beam is shown in Fig. 1(d).The bending moment was applied to these cantilevers using Nano-XP platform (Agilent) equipped with a sphero-conical diamond indenter (1 μm diameter tip radius). A constant displacement rate of 5 nm/s was applied until the beams fractured, while the load, displacement, and harmonic contact stiffness were recorded. Fracture events resulted in large displacement jumps and fracture was defined when the displacement changed by more than 100 nm between data points, compared with the 1-5 nm displacement steps between data points that was typical during loading.

Table 1.Microstructural features of the Al/SiC nanolaminates quantified inside the TEM.

| Nanolaminate | Thickness μm | Layer Thickness | | Al grain size | | Columnar Width, $d_c$ nm |
|---|---|---|---|---|---|---|
| | | $h_{Al}$ (nm) | $h_{SiC}$ (nm) | vertical, $d_v$ (nm) | lateral, $d_l$ (nm) | |
| AS10 | ~12 | 8±1 | 11±2 | 8±1 | 29±6 | 171±27 |
| AS25 | ~14 | 25±4 | 25±7 | 25±4 | 47±11 | 305±73 |
| AS50 | ~15 | 52±2 | 44±2 | 52±2 | 48±7 | 450±105 |
| AS100 | ~17 | 100±6 | 148±5 | 100±6 | 62±12 | 713±221 |

## 3. Results and discussion

### 3.1.Micropillar splitting

The force-displacement curves of Al/SiC nanolaminates obtained in the micropillar splitting tests are shown in Fig. 2. Up to 5 tests were carried out for each layer thickness and the curves show good reproducibility. They are characterized by a monotonous increase of the load up to a critical load ($P_c$) that indicates the micropillar splitting. It is clear that the critical load increased considerably as the layer thickness decreased from 100 nm to 25 nm, and dropped again for a layer thickness of 10 nm. One interesting observation is the more compliant behavior of the AS10 nanolaminate, compared to the rest of nanolaminates. It is speculated, based on the observation of the crack paths that will be presented below, that this extra compliance could come from the columnar boundaries. As shown in Fig. 1(b), the AS10 nanolaminate contained a higher density of columnar boundaries. Interestingly, the micropillars with layer thickness of 100 nm and 50 nm were able to sustain some load after the onset of splitting, which indicates that the splitting process was gradual. The micropillars from AS25 and AS10 failed in a more brittle fashion, as indicated by the arrows.

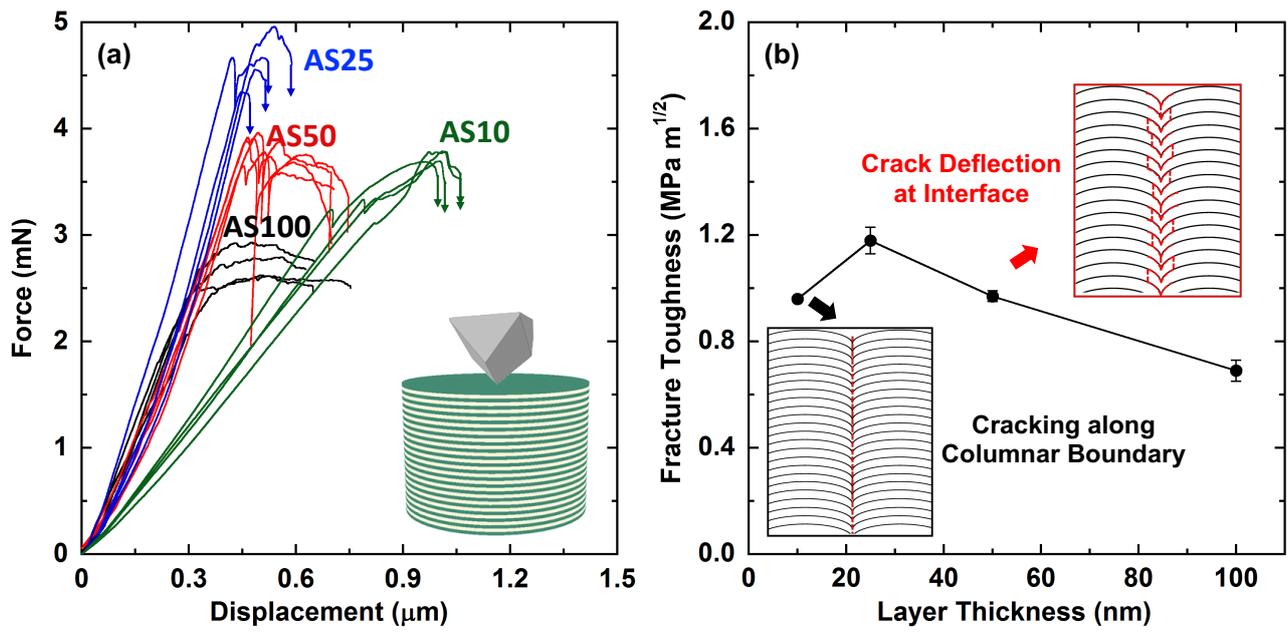

Fig. 2. (a) Force-displacement curves from the micropillar splitting tests of Al/SiC nanolaminates with different layer thickness; (b) Fracture toughness (measured with the micropillar splitting test) vs. layer thickness for Al/SiC nanolaminates.

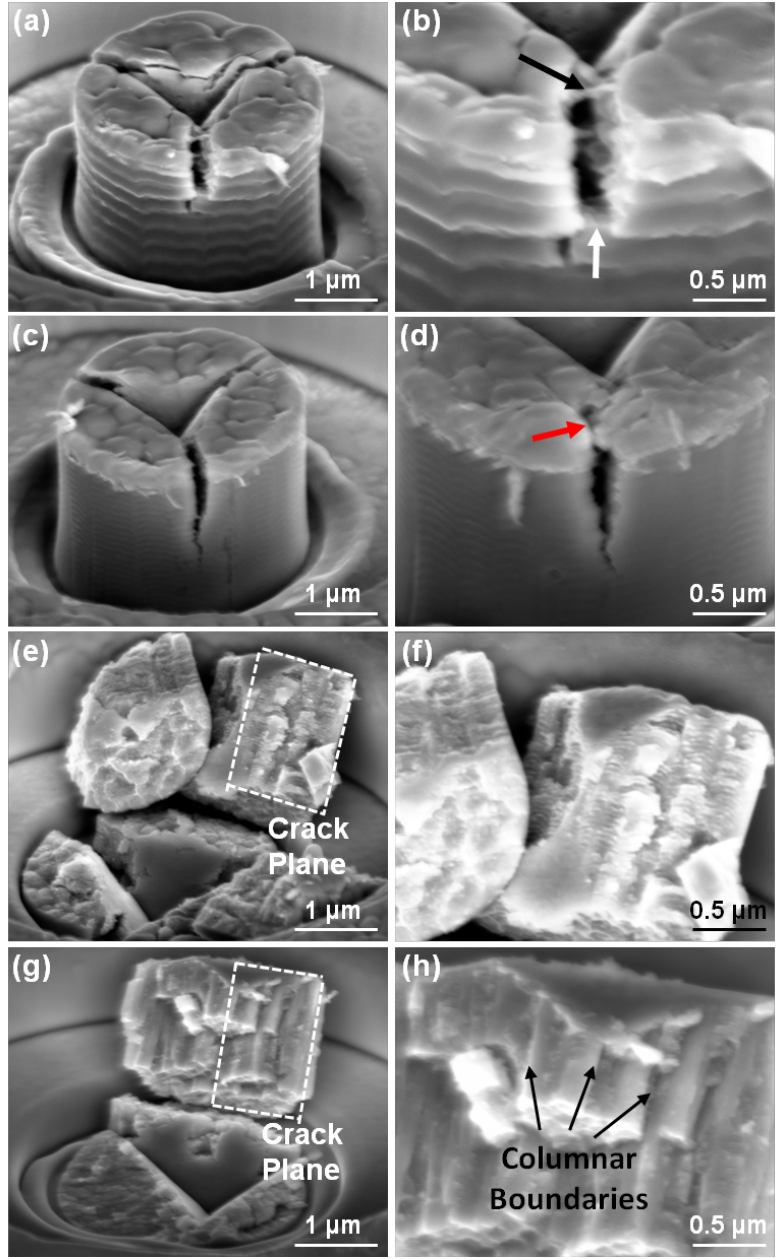

Fig. 3. SEM images of the micropillars after the splitting tests for (a,b) AS100 nanolaminate, (c,d) AS50 nanolaminate, (e,f) AS25 nanolaminate and (g,h) AS10 nanolaminate.

The fracture toughness of each nanolaminate, $K_{IC}$, was determined from the splitting load, $P_C$, according to [20-21]

$$K_{IC} = \gamma \frac{P_C}{R^{3/2}} \qquad (1)$$

where $R$ is the micropillar radius and $\gamma$ a parameter that depends on the micropillar volume, the elasto-plastic properties and the indenter geometry. This parameter was determined from the finite element simulations presented in the supplementary material and was in the range 0.46-0.47 for the four nanolaminates.

The experimental results of the fracture toughness are plotted as a function of the nanolaminate layer thickness in Fig. 2(b). It is low in all cases, of the order of 0.7-1.2 MPa √m, probably due to the limited plasticity of the Al layers at these length scales and to the brittleness of the ceramic SiC layers.

Interestingly, the fracture toughness increased when the layer thickness decreased from 100 nm to 25 nm, and decreased again for the smallest layer thickness of 10 nm. It is expected that, in the case of metal-ceramic laminates with cracks propagating perpendicular to the layers, the stresses at the crack tip might be relaxed by mechanisms such as the plastic deformation of the metallic layers as well as crack arrest and crack deflection at the metal-ceramic interfaces [9]. In this case, some ligaments could be seen in the fracture surfaces of Fig. 3(b) (black arrows), which presumably indicate that the Al layers may experience ductile fracture. However, it is unlikely that plasticity in the Al layers is responsible for the increase in fracture toughness with layer thickness reduction because the yield stress of the Al layers increases with layer thickness reduction from 100 nm to 25 nm (from 900 to 1200 MPa [22]) and the size of the plastic zone ahead of the crack tip may be hampered by the presence of the interfaces. As a matter of fact, the fracture toughness of thin Al films with thickness of 100-125 nm (grain size ~50 nm) has been reported to be rather small, 0.7~1.1 MPa √m [24], which is one order of magnitude lower than that for pure bulk Al.

The fracture surfaces in Fig. 3(a) to 3(f) showed that the cracks seemed to propagate close to the pre-existing columnar boundaries for layer thicknesses of 100 nm (Figs. 3(a) and (b)), 50 nm (Figs. 3(c) and (d)) and 25 nm (Figs. 3(e) and )f)), and they followed a tortuous path with clear signs of crack deflection (white arrows) at the Al-SiC interfaces. A schematic illustration of the interaction of the cracks with interfaces and columnar boundaries in this case is inserted in Fig. 2(b). Considering that the interface density increases with $1/h$, it is reasonable to assume that the increase in fracture toughness with layer thickness reduction may be a consequence of the higher energy dissipation by crack arrest/crack deflection at interfaces. Additionally, if cracks are arrested at the Al-SiC interfaces, higher fracture stresses are expected for the thinner SiC layers.

The fracture toughness was reduced, however, to ~0.96 MPa √m for the thinnest AS10 nanolaminates and its fracture surface was different from the others, as seen in Figs. 3(g) and 3(h). In this case, the crack propagated leaving behind smooth terraces that do not show any traces of the energy dissipation mechanisms found for the thicker layers (crack deflection at the interfaces or plasticity of the Al layers). Moreover, the morphology of the terraces indicates that the crack propagation took place along the pre-existing columnar boundaries. A schematic illustration of the interaction of the cracks with interfaces and columnar boundaries in this case is also inserted in Fig. 2(b). It is possible that plasticity in the Al layers is hampered for such thin layers, preventing the cracks to be arrested at interfaces. However, it should be reminded that TEM observations revealed extremely narrow columnar grains in the AS10 nanolaminate, and that there were doubts about the continuity of the layers across the columnar boundaries. This may lead to the propagation of the cracks along the weaker columnar boundaries, and to the drop observed in the fracture toughness. In fact, there are other observations that suggest that the AS10 nanolaminate might be more influenced by the columnar boundaries than the rest of the nanolaminates, such as the drop in hardness, as reported in [22], the more compliant behavior of the load-displacement curves in the micropillar splitting tests (Fig. 2) or the cracks generated along columnar boundaries in the micropillar compression experiments [3].

*3.2. Fracture of notched cantilevers*

The load-displacement curves corresponding to the bending test in the notched cantilevers are plotted in Figs. 4(a) and 4(b) for nanolaminates with the layers perpendicular and parallel to the notch plane, respectively. Qualitatively, the load-displacement curves are approximately linear for the perpendicular case, which indicates there is not a significant amount of stable crack growth or plasticity at the crack tip. Nevertheless, the curves of the cantilevers deformed in parallel orientation presented higher nonlinearity near to the maximum load as the layer thickness increased, and this behavior could be associated with larger contributions of the plastic deformation to the energy dissipation near the crack tip.

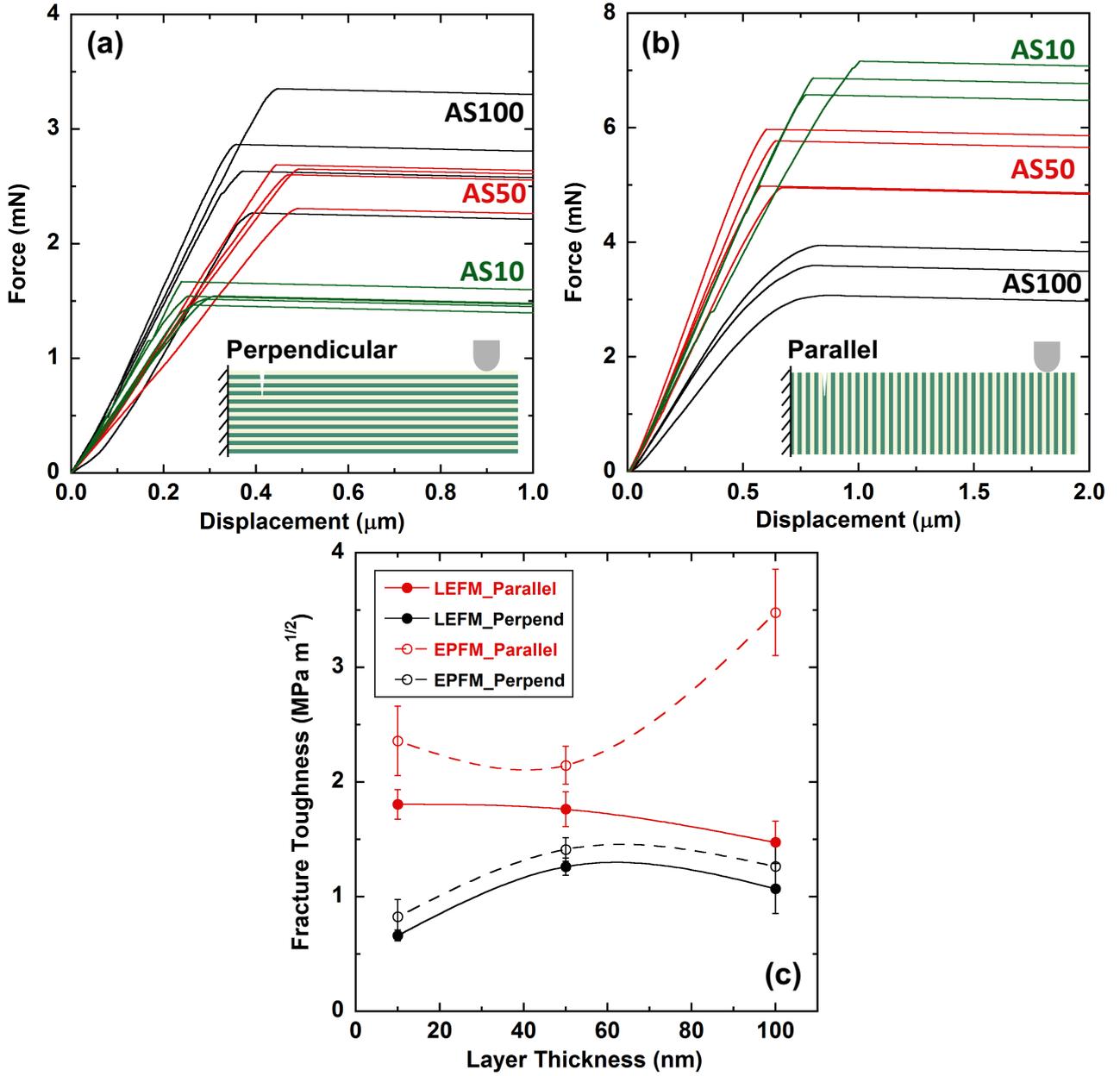

Fig. 4. Load-displacement curves corresponding to the bending test in the notched cantilevers: (a) Layers perpendicular to the notch plane; (b) Layers parallel to the notch plane; (c) Fracture toughness calculated according to linear elastic and elasto-plastic fracture mechanics principles.

The linear elastic fracture mechanics approach to determining the toughness of materials through cantilever beams has been widely used in the literature [15, 24]. The fracture toughness is calculated using the peak load applied to the beam ($P_{max}$), a dimensionless constant based on the sample geometry ($f_{CB}$), along with the beam dimensions ($L$, length; $W$, width; $B$, thickness; $a$, depth of pre-notch), according to [15, 25]

$$K_{LEFM} = \frac{P_{max} L}{B W^{1.5}} f_{CB} \quad (2)$$

$$f_{CB} = 1.46 + 24.36(\frac{a}{W}) - 47.21(\frac{a}{W})^2 + 75.18(\frac{a}{W})^3 \quad (3)$$

The variation in measured toughness as a function of layer thickness and orientation are plotted in Fig. 4(c). The most widely accepted criterion for determining whether the size of the specimen is adequate to achieve plane strain conditions at the crack tip is given by $B, a \geqslant 2.5(K_{LEFM}/\sigma_y)^2$. From the toughness values in Fig. 4(c) and the yield strength from previous indentation studies [26], the $B$ and $a$ dimensions would need to be approximately 1.5 μm. As the notch length in the cantilever beams was smaller than this magnitude, these tests do not satisfy the plane strain condition. This shows that it can be challenging to obtain valid plane strain fracture toughness measurements at the microscale even in materials showing fairly brittle behaviour because of the small sample dimensions.

Nevertheless, these limitations can be overcome through the application of elasto-plastic fracture mechanics by means of the $J$ integral. The $J$ integral is a measure of the amount of energy required to propagate a crack, identical to the strain energy release rate, $G$, in the case of elastic materials [27]. As such, the elasto-plastic fracture toughness ($K_{JC}$) can be determined from $J$ according to:

$$K_{JC} = \sqrt{\frac{JE}{1-v^2}} \qquad (4)$$

where $E$ and $v$ stand for the elastic modulus and the Poisson's ratio of the material, which is assumed to be isotropic. They were chosen based on nanoindentation results [28] and can be found in Table S1 in the supplementary material for each nanolaminate.

The $J$ integral includes two independent contributions from the elastic ($J_{el}$) and the plastic energy ($J_{pl}$) dissipated during fracture,

$$J = J_{el} + J_{pl} \qquad (5)$$

The elastic contribution is calculated from the fracture toughness determined using linear elastic fracture mechanics (eq. (2)),

$$J_{el} = \frac{K_{LEFM}^2 (1-v^2)}{E} \qquad (6)$$

while $J_{pl}$ can be determined from the load-displacement curves following the methodology detailed in the supplementary material.

The elasto-plastic fracture toughness, $K_{JC}$, calculate from eq. (4) is also plotted in Fig. 4(c) for each nanolaminate in the perpendicular and parallel orientations and it should be noted that all the toughness values obtained from the $J$ integral fulfil the validity criterion of $B, (W-a) \geqslant 25J/\sigma_y$ for plane strain fracture toughness [27]. The differences with the fracture toughness determined from linear elastic fracture mechanics were very small in the perpendicular orientation as well as for the AS10 and AS50 nanolaminates in the parallel orientation. Nevertheless, the elasto-plastic fracture toughness of the AS100 nanolaminate in the parallel orientation was much higher than the elastic one, indicating that the contribution of plastic deformation was significant in this case. This behavior is in agreement with previous micropillar compression tests which showed that plastic deformation of the Al layers was inhibited in the case of the AS10, AS25 and AS50 nanolaminates due to the constrain imposed by the SiC layers and by the small Al grain size. Nevertheless, plastic flow of the Al, which was extruded from the lateral micropillar surface during micropillar compression, was observed in the AS100 nanolaminates [29].

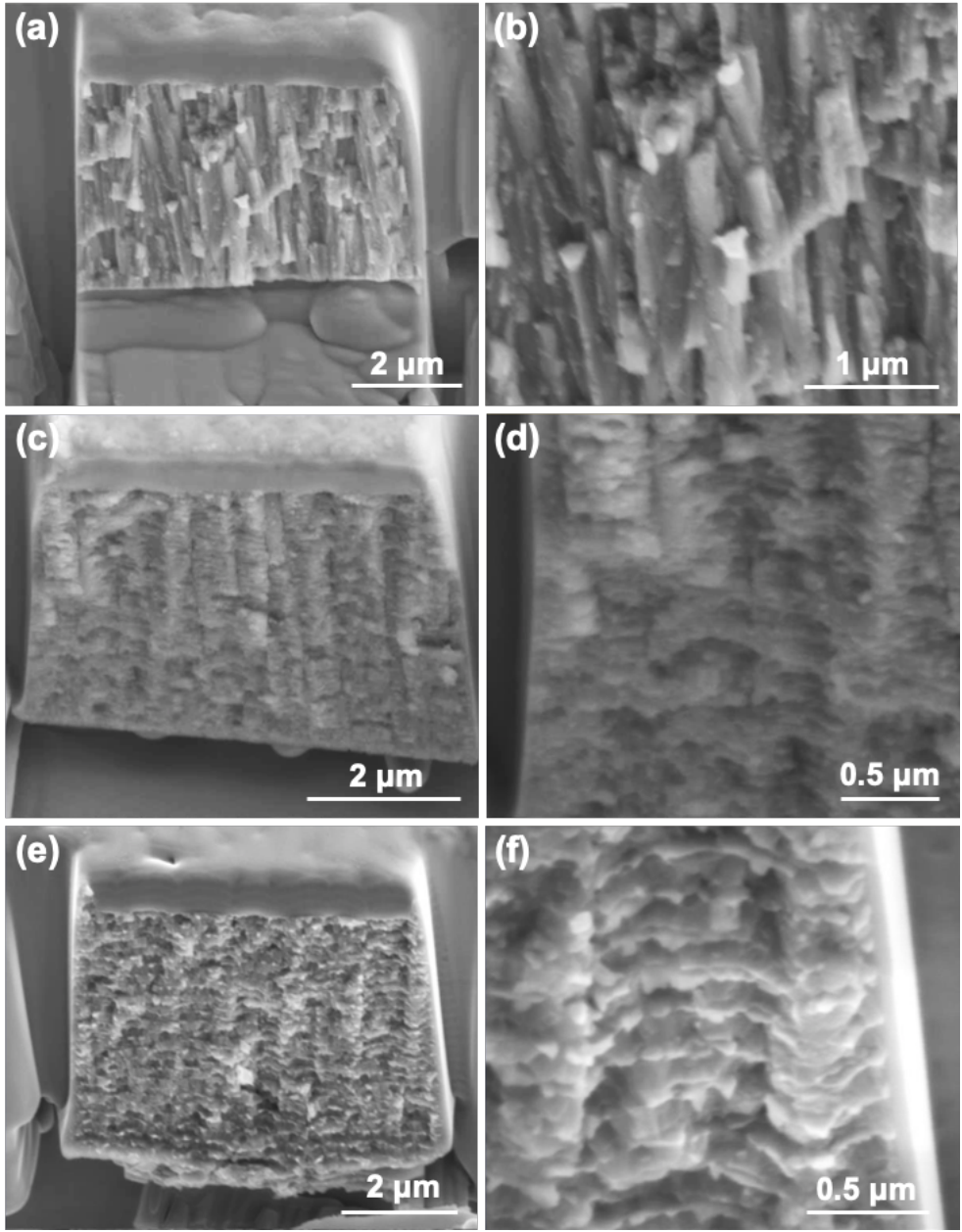

Fig. 5. SEM images of the fracture surfaces of the perpendicular oriented beams: (a-b) AS10 nanolaminate; (c-d) AS50 nanolaminate; (e-f) AS100 nanolaminate.

Insights on the fracture mechanisms in the notched cantilevers can be obtained from the SEM micrographs of the fracture surfaces in the perpendicular (Fig. 5) and parallel (Fig. 6) orientations. In the case of the perpendicular orientation, the AS10 specimen did not show any horizontal striations which would be indicative of crack blunting due to the Al layers and the fracture path appeared to be dominated by the vertical columnar grain boundaries (Figs. 5a and 5b). The AS50 and AS100 specimens (Figs. 5c to 5f) showed significant horizontal striations, which were indicative of crack arrest/crack deflection at interfaces. These observations are in agreement with the results obtained in the micropillar splitting tests. As a matter of fact, it is remarkable that, despite the small quantitative differences in the actual toughness values obtained from micropillar splitting tests and the micro-cantilever bending tests in the perpendicular orientation, both show the same trend with layer thickness.

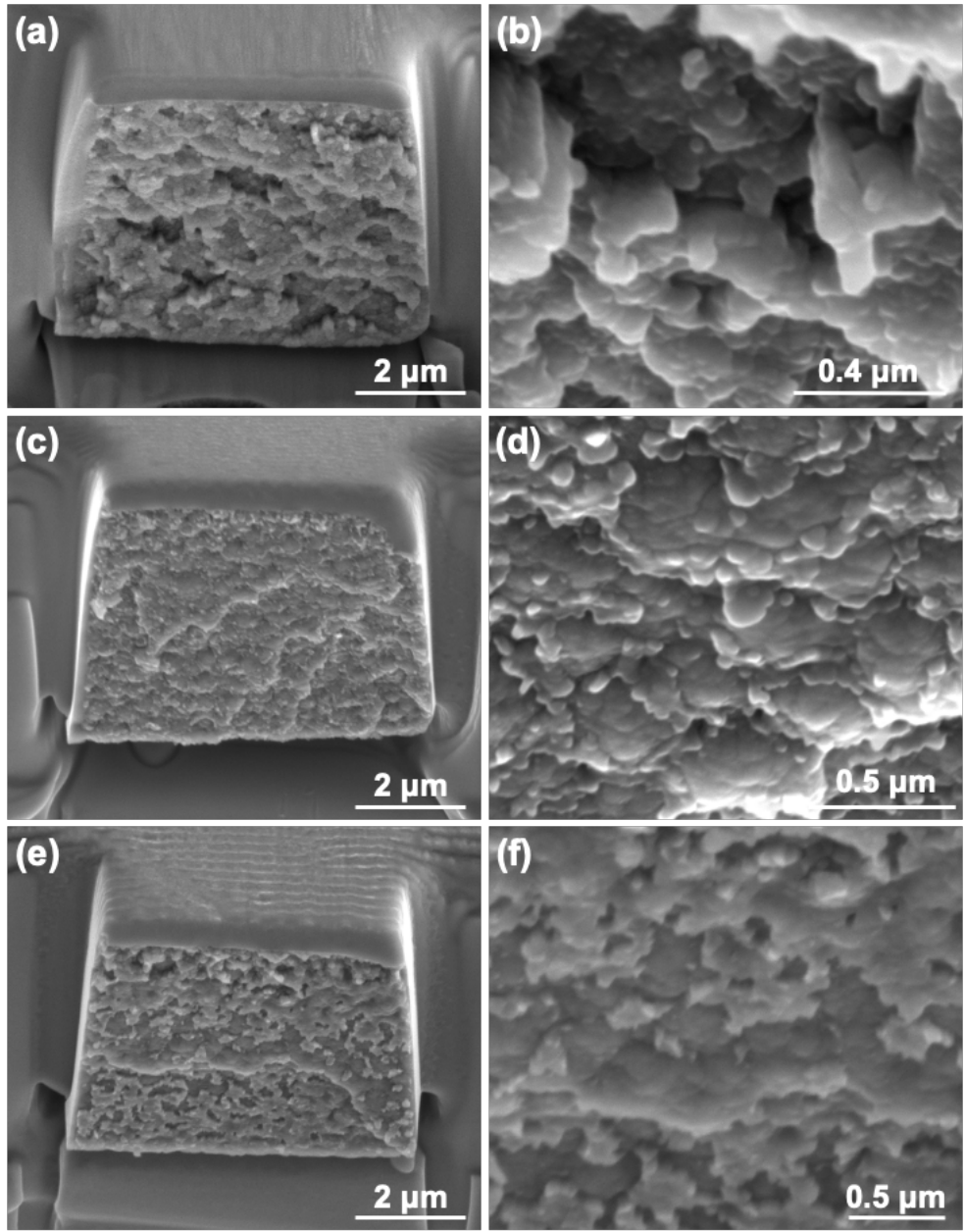

Fig.6. SEM images of the fracture surfaces of the parallel oriented beams: (a-b) AS10 nanolaminate; (c-d) AS50 nanolaminate; (e-f) AS100 nanolaminate.

The fracture toughness of the nanolaminates in the parallel orientation was higher than that in the perpendicular orientation, particularly for the AS10 nanolaminate. The magnitude of the roughness in the fracture surface of the AS10 nanolaminate was much larger than the layer dimensions (Figs. 6a and 6b), indicating that the fracture path was not through a single layer or interface but rather it propagated through multiple layers. In the case of the AS50 and AS100 nanolaminates, the fracture surfaces in Figs. 6c to 6f showed the presence of discrete islands of Al remaining on the surface. The height of these islands was approximately equal to the individual layer thickness in each nanolaminate, indicating that crack propagation took predominately place along the interface with occasional fractures through the Al layers to reach the adjacent interface.

Overall, the toughness of the nanolaminates in the parallel orientation was higher than that in the perpendicular orientation and the largest differences were found in the AS10 and AS100 nanolaminates. The effect of orientation in the AS10 nanolaminate may be attributed to the presence of the columnar grain boundaries, which provided an easy path for crack propagation in the perpendicular orientation. In case of the AS100 nanolaminate, the differences between the parallel and perpendicular orientations can be attributed to the contribution of the plastic deformation of the Al layers to the toughness in the former orientation.

## 4. Conclusions

The fracture toughness of Al/SiC nanolaminates with different layer thicknesses (in the range 10 to 100 nm) was measured by means of two different micromechanical testing techniques, micropillar splitting and bending of a notched beam. The crack path was perpendicular to the layers in the former and parallel and perpendicular to the crack plane in the latter. Overall, the toughness of the nanolaminates in the parallel orientation was higher than that in the perpendicular orientation. In the parallel orientation, the cracks propagated parallel to the interface in the nanolaminates with 50 and 100 nm layer thickness and through multiple layers in the nanolaminates with 10 nm layer thickness. The toughness of the nanolaminate in this orientation increased with the Al layer thickness to reach a maximum for 100 nm layer thickness, due to the contribution of the plastic deformation of the Al layers. In the perpendicular orientation, the fracture surfaces showed horizontal striations, which were indicative of crack arrest/crack deflection at interfaces. The toughness increased as the layer thickness decreased in this orientation up to a maximum for the nanolaminate with 25 nm layer thickness. Nevertheless, the toughness of the nanolaminate with 10 nm layer thickness in the perpendicular orientation was significantly lower because of the preferential propagation of the crack along the weak columnar grain boundaries.

**Acknowledgements**

This investigation was supported by the European Research Council (ERC) under the European Union's Horizon 2020 research and innovation programme (Advanced Grant VIRMETAL, grant agreement No. 669141). Additional support is gratefully acknowledged from the U.S. National Science Foundation and the Spanish Ministry of Economy and Competitiveness under the Materials World Network Program through the project "High temperature mechanical behavior of metal/ceramic nanolaminate composites" (Dr. Lynnette Madsen, NSF-DMR-1209988, PCIN-2013-029 and MAT2012-31889).

# SUPPLEMENTARY MATERIAL

## 1. Determination of the fracture toughness from the micropillar splitting tests

The determination of the fracture toughness from the critical splitting load relied on the finite element simulation of the micropillar splitting test to determine the constant $\gamma$ in eq. (1). Neither the layered structure nor the columnar boundaries were considered explicitly in the model, i.e., the nanolaminates were modelled as homogeneous materials with the elastic and plastic properties of the corresponding nanolaminate. Fig. S1 shows the geometry of the finite element model. Only 1/3 of the micropillar and of the rigid cube-corner indenter were modelled because of the symmetry. Symmetric boundary conditions were imposed on the left and right surfaces of the micropillar. The crack plane, aligned with the corner of the indenter and indicated in red in Fig. S1, was introduced with cohesive surfaces. A total number of ~20,000 8-node linear brick (C3D8) elements were used to mesh the micropillar, with mesh sizes that were refined in the area of initial contact with the indenter. The elastoplastic properties of the homogeneous materials representing each nanolaminate were chosen based on nanoindentation results [S1], assuming a Tabor factor of ~3, and they are depicted in Table S1.

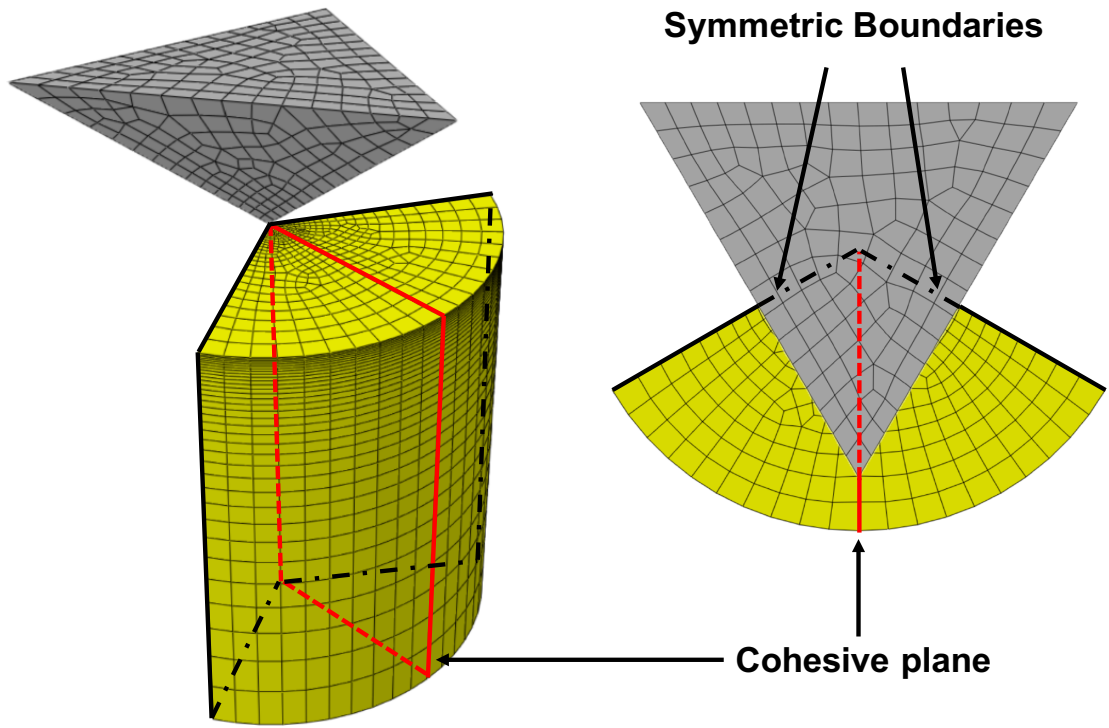

Fig. S1. Finite element model of a micropillar splitting test.

Table S1. Elastic modulus ($E$), Poisson's ratio ($\nu$), hardness ($H$) and yield strength ($\sigma_y$) of the homogeneous materials representing each nanolaminate in the finite element simulation of the micropillar splitting tests.

| Nanolaminate | $E$ (GPa) | $\nu$ | $H$ (GPa) | $\sigma_y$ (GPa) |
|---|---|---|---|---|
| AS100 | 140 | 0.2 | 5.7 | 2.11 |
| AS50 | 141 | 0.2 | 5.7 | 2.11 |
| AS25 | 141 | 0.2 | 6.8 | 2.52 |
| AS10 | 140 | 0.2 | 6.5 | 2.41 |

The maximum normal stress criterion was used to model the cohesive fracture, using the traction-separation law plotted in Fig. S2. The initial stiffness, $S$ and cohesive strength, $\sigma_{max}$, of the crack were set to $10^5$ GPa and 500 MPa, respectively in all simulations. The initial stiffness is much greater than the elastic modulus to minimize artificial compliances introduced by the cohesive elements. Moreover, an artificial viscosity of $10^{-6}$ s$^{-1}$ was used to aid convergence during softening/debonding by forcing a positive tangent stiffness matrix over a short period of time. These values were necessary to satisfy both convergence and the criteria for brittle behavior. Two different values of the critical energy release rate (fracture energy), $G_c$, determined from the area beneath the traction-separation curve, were used in the simulations (1 J/cm$^2$, 0.5 J/cm$^2$) to ascertain load start to drop, indicating the fracture of the cohesive surface.

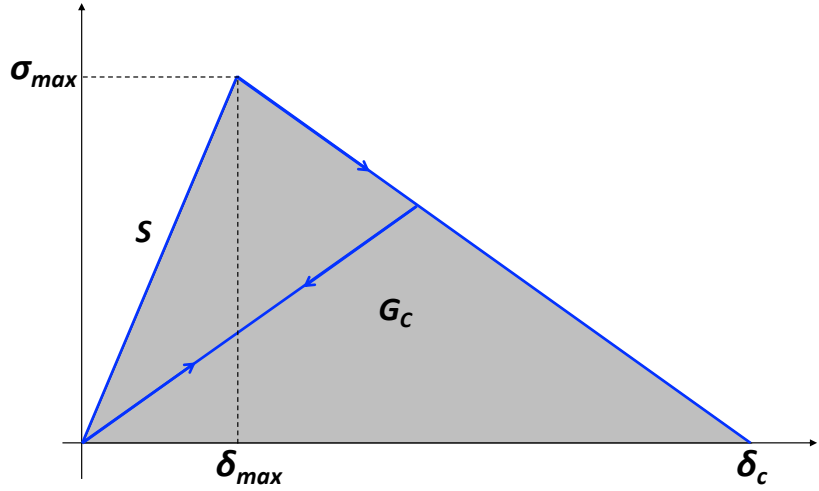

Fig. S2. Constitutive traction-separation relation for the cohesive interface crack, where $S$ is the initial stiffness, and $\sigma_{max}$ the cohesive strength. The area under the traction-separation curve determines the fracture energy $G_c$.

The non-dimensional load, defined as $P_c/K_{IC}R^{3/2}$, is plotted in Fig. S3 as a function of the non-dimensional displacement into surface, $h/R$, (where $h$ is penetration depth) for the micropillars of the different nanolaminates. The normalized load-displacement behavior resembles the elastoplastic indentation behavior of bulk materials until a sharp load drop occurs at a critical point. This behavior is consistent with the experimental observations (Fig. 2a). Note that the curves in Fig. S3 indicate that the non-dimensional load drop occurs consistently at an approximately constant value for a given material, independent of the fracture energy, $G_c$.

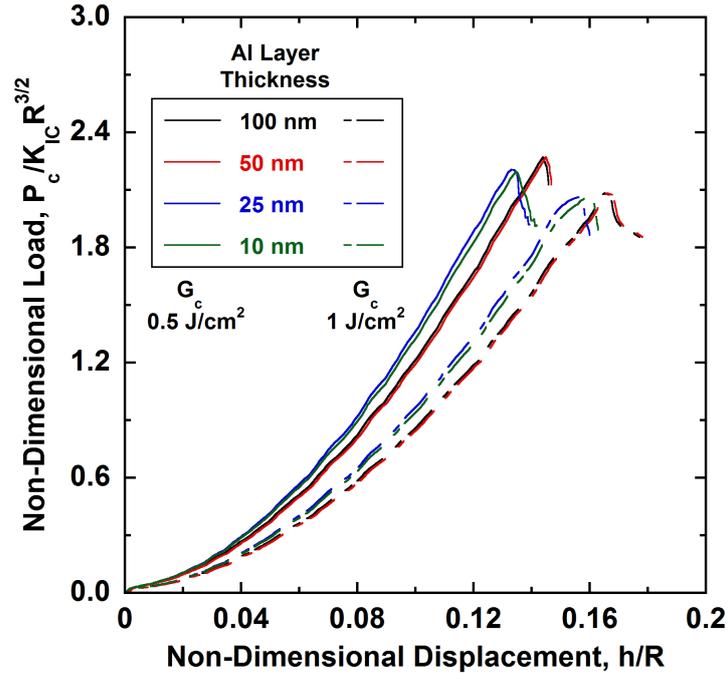

Fig. S3. Non-dimensional load as a function of the non-dimensional displacement into surface for the Al/SiC micropillars with different layer thickness.

Initial contact of the cube-cornered indenter with the pillar resulted in elastic-plastic deformation with a zone of plastically deformed material near the contact that scaled in size with the applied displacement. A crack nucleated outside of the plastic zone according to the constitutive behavior of the cohesive elements. Median/radial cracks developed with stable growth occurring both at the surface and subsurface as displacement increased. As the crack grew, an unstable crack propagation occurred when the maximum load was reached, leading to the micropillar splitting, as shown in Fig. S4.

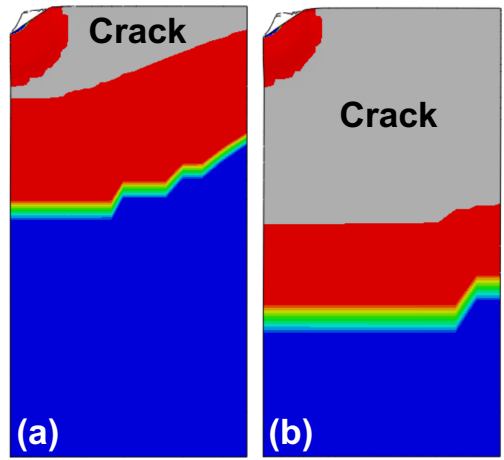

Fig. S4. Unstable crack propagation at: (a) one step before and (b) one step after the maximum load in the finite element simulations.

The simulation results were then used to calibrate $\gamma$ in eq. (1), which only depends on the material properties ($E/H$) and indenter geometry. $G_c$ is related to the fracture toughness, $K_{IC}$, through linear elastic fracture mechanics according to

$$K_{IC} = \sqrt{\frac{EG_c}{1-\nu^2}} \tag{S1}$$

Once $K_{IC}$ was estimated from the simulations, γ is given as:

$$\gamma = \frac{K_{IC} R^{3/2}}{P_{c\_FEM}} \quad (S2)$$

where $P_{C\_FEM}$ is the maximum load in the finite element simulations. The constant γ was 0.46 for the AS100 and AS50 nanolaminates and 0.47 for the AS25 and AS10 nanolaminates.

**2. Determination of the fracture toughness from the *J* integral in the notched cantilever beams**

The plastic contribution to the J integral is given by [S2]

$$J_{pl} = \frac{\eta A_{pl}}{B(W-a)} \quad (S3)$$

where $A_{pl}$ stands for energy dissipated due to plastic deformation and $\eta$ a geometric factor that depends on the sample configuration. $A_{pl}$ can be calculated from the total area under the load-displacement curve up to the point of maximum load in Figs. 4a and 4b minus the elastic energy stored in specimen, which is determined assuming the elastic stiffness during unloading was parallel to the elastic stiffness during loading. As it will be shown below, stable crack propagation did not take place during loading and, thus, *a* stands for the initial notch length in eq. (S3).

The geometric factor $\eta$ has been determined for a notched beam subjected to three-point bending [S2], and it is expressed as

$$\eta_{3pt} = 2 - \frac{(1-\frac{a}{W})(1.3096 - 1.6314\frac{a}{W})}{0.9534 + 1.3096\frac{a}{W} - 0.8157(\frac{a}{W})^2} \quad (S4)$$

A notched beam with span *S* subjected to three-point bending is geometrically equivalent to a cantilever beam with length $L = S/2$ and the only difference is that the load measured at the center of the three-point bend beam is twice larger than the one measured at the tip of the cantilever. So, the actual geometric coefficient for the cantilever, $\eta_C$, can be approximated by $\eta_{CB} \cong 2\eta_{3pt}$.

Finally, estimation of the plastic contribution to the toughness using the *J* integral requires to account for any change in crack length due to stable crack growth during the test [S2]. The harmonic contact stiffness vs displacement curves (Fig. S5) were used to determine if any crack growth took place prior to the maximum load in the test, which was followed by the sudden propagation of the crack until failure. The measured harmonic contact stiffness is a combination of both the cantilever beam stiffness and the surface contact stiffness between the indenter and the surface, which behave as two springs in series. The surface contact stiffness increases sharply with load at low displacements to become constant while the beam stiffness is constant. Any stable crack growth near the maximum load (which corresponds to displacements > 400 nm in the parallel orientation and > 200 nm in the perpendicular orientation) should result in a significant decrease in the beam stiffness, which is not observed in any of the curves. These results indicate that there is no crack growth and that *J* can be calculated using the initial crack length without any additional corrections.

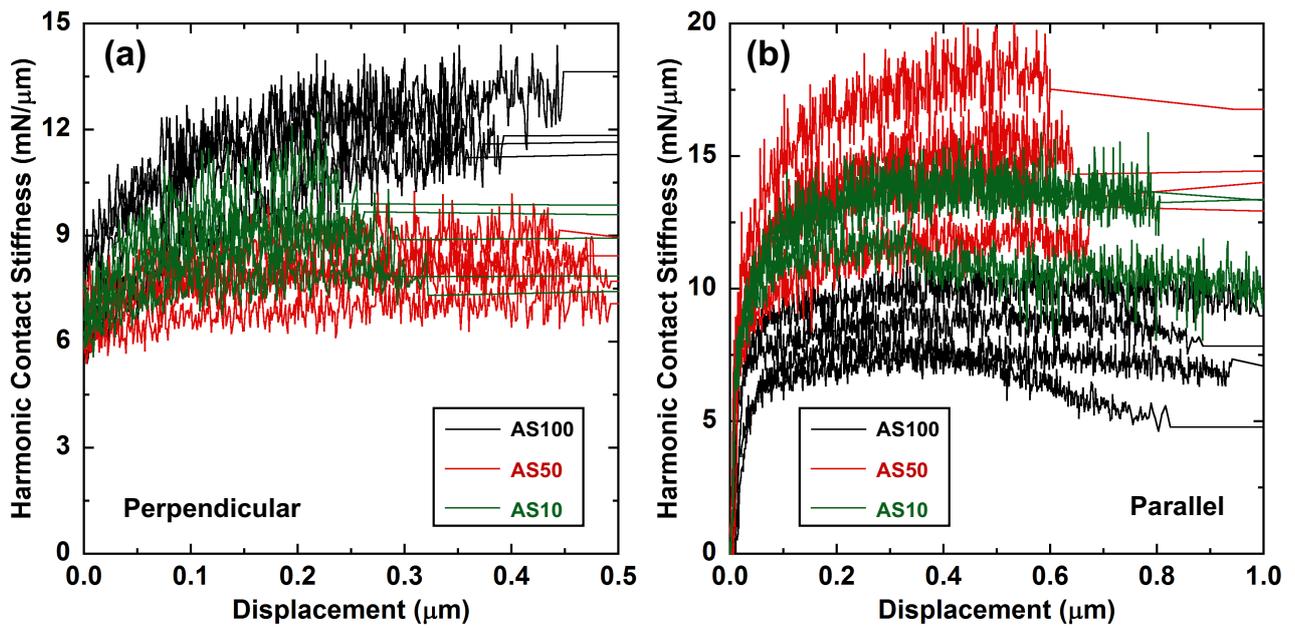

Fig. S5. Evolution of the harmonic contact stiffness as a function of the displacement for each cantilever beam. (a) Perpendicular orientation. (b) Parallel orientation.

**References**

[S1] S. Lotfian, J.M. Molina-Aldareguia, K.E. Yazzie, J. Llorca, N. Chawla, High-temperature nanoindentation behavior of Al/SiC multilayers, Philos. Mag. Lett. 92 (2012) 362–367.

[S2] X.-K. Zhu and J. A. Joyce, Review of fracture toughness (G, K, J, CTOD, CTOA) testing and standardization, Engng.Frac. Mech. 85 (2012) 1–46.